\documentclass[sigconf]{acmart}
\usepackage{booktabs} 

\newcommand{\ignore}[1]{}
\newcommand{\an}[1]{$^#1$}

\copyrightyear{2019}
\acmYear{2019}
\setcopyright{acmlicensed}
\acmConference[SIGIR '19] {42nd International ACM SIGIR Conference on Research and Development in Information Retrieval}{July 21--25, 2019}{Paris, France}
\acmBooktitle{42nd Int'l ACM SIGIR Conference on Research and Development in Information Retrieval (\mbox{SIGIR '19}), July 21--25, 2019, Paris, France}
\acmPrice{15.00}
\acmDOI{10.1145/3331184.3331340}
\acmISBN{978-1-4503-6172-9/19/07}

\begin{document}
\title[Critically Examining the ``Neural Hype'']{Critically Examining the ``Neural Hype'': Weak Baselines and the Additivity of Effectiveness Gains
from Neural Ranking Models}

\author{Wei Yang,\an{1} Kuang Lu,\an{2} Peilin Yang, and Jimmy Lin\an{1}}
\affiliation{\vspace{0.1cm}
  \institution{$^{1}$ David R. Cheriton School of Computer Science, University of Waterloo}
  \institution{$^{2}$ Department of Electrical and Computer Engineering, University of Delaware}
}

\renewcommand{\shortauthors}{Wei Yang, Kuang Lu, Peilin Yang, and Jimmy Lin}

\begin{abstract}
Is neural IR mostly hype?
In a recent SIGIR Forum article, Lin expressed skepticism that neural ranking models were actually improving {\it ad hoc} retrieval effectiveness in limited data scenarios.
He provided anecdotal evidence that authors of neural IR papers demonstrate ``wins'' by comparing against weak baselines.
This paper provides a rigorous evaluation of those claims in two ways:
First, we conducted a meta-analysis of papers that have reported experimental results on the TREC Robust04 test collection.
We do not find evidence of an upward trend in effectiveness over time.
In fact, the best reported results are from a decade ago and no recent neural approach comes close.
Second, we applied five recent neural models to rerank the strong baselines that Lin used to make his arguments.
A significant improvement was observed for one of the models, demonstrating additivity in gains.
While there appears to be merit to neural IR approaches, at least some of the gains reported in the literature appear illusory.
\end{abstract}

\maketitle

\fancyhead{}

\section{Introduction}

In a recent SIGIR Forum opinion piece, Lin~\cite{Lin_SIGIRForum2018} criticized the state of information retrieval research, making two main points.
First, he lamented the ``neural hype'' and wondered that for ``classic'' {\it ad hoc} retrieval problems (limited relevance judgments and no behavioral data), whether neural ranking techniques represented genuine advances in effectiveness.
As anecdotal evidence, he discussed two recent papers that demonstrated improvements over weak baselines, but in absolute terms, the reported results were no better than a well-tuned bag-of-words query expansion baseline.

In this paper, we attempt a rigorous evaluation of these claims.
Focusing specifically on the test collection from the TREC 2004 Robust Track, a meta-analysis of the literature shows no upward trend in reported effectiveness over time.
The best reported results on the collection are from a decade ago, and no recent paper (using neural approaches or otherwise) has reported effectiveness close to those levels.
Analysis of over one hundred papers confirms that the baseline comparison conditions are often not as strong as they should be.
Thus, Lin's criticism that comparisons to weak baselines still pervade the IR community rings true.

As a follow up, we applied a number of recent neural ranking models from the MatchZoo toolkit~\cite{fan2017matchzoo} to rerank the strong baselines that Lin used to make his arguments.
Out of five neural models, one was able to significantly improve upon Lin's results.
In other words, the effectiveness gains from one neural model is additive with respect to a strong baseline---which provides evidence that neural IR {\it can} lead to ``real'' improvements.
Nevertheless, four out of the five models examined were not able to significantly beat the baseline, suggesting that gains attributable to neural approaches are not as widespread as the literature suggests.
The absolute average precision values we report are among the highest for neural models that we are aware of, although in absolute terms they are still much lower than the best known results.

\section{Meta-Analysis}

\ignore{
\begin{table}[t]
\centering
\begin{tabular}{lrl}
\toprule
Year & Count & Venues \\
\toprule
2005 & 3  & SIGIR (1), CIKM (2) \\
2006 & 3  & SIGIR (2), ECIR (1) \\
2007 & 4  & SIGIR (1), CIKM (1), TOIS (1) , IRJ (1) \\
2008 & 6  & SIGIR (4), ECIR(1), IPM (1) \\
2009 & 12 & SIGIR (7), CIKM (1), ECIR (1), ICTIR (2), IPM (1) \\
2010 & 9  & SIGIR (3), CIKM (4), WSDM (1), ECIR (1) \\
2011 & 8  & SIGIR (2), CIKM (4), ICTIR (1), IRJ (1) \\
2012 & 6  & SIGIR (2), CIKM (1), WSDM (1), ECIR (1), IRJ (1) \\
2013 & 6  & SIGIR (1), CIKM (1), ICTIR (1), TOIS (1), IPM (2) \\
2014 & 6  & SIGIR (2), TOIS (1), IRJ (2), JASIST (1) \\
2015 & 6  & SIGIR (1), CIKM (1), ICTIR (2), TOIS (1), IRJ (1) \\
2016 & 18 & SIGIR (3), CIKM (7), ICTIR (3), WWW (1),\\
     &    & ECIR (2), TOIS (1), JASIST (1)\\
2017 & 9 & SIGIR (5), WSDM (1), ECIR (1), ICTIR (2)\\
2018 & 12 & SIGIR (2), CIKM (3), ECIR (2), ICTIR (2),\\
     &    & IPM (1), EMNLP (2)\\
\bottomrule
\end{tabular}
\caption{Summary of papers examined.}\label{papers}
\vspace{-0.75cm}
\end{table}
}

\begin{figure*}[t]
\centering
\includegraphics[width=1.0\textwidth]{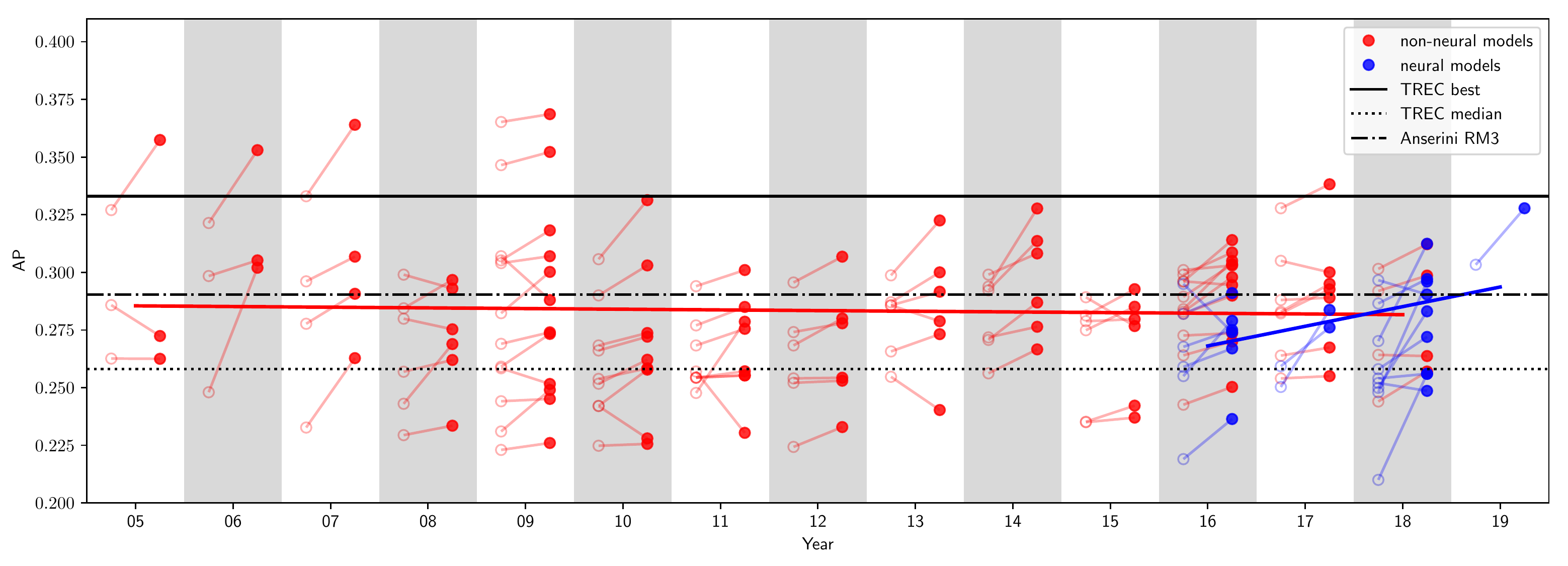}
\vspace{-0.8cm}
\caption{Visualization results on Robust04, where baseline and best AP scores are represented by empty and filled circles.}
\label{fig:scatter}
\vspace{-0.1cm}
\end{figure*}

The broader context of Lin's article is a recent series of papers that reflects a general angst (at least by some researchers) about the state of machine learning and its applications, in particular regarding empirical rigor and whether genuine advances are being made~\cite{Sculley_etal_2018,Lipton:1807.03341v2:2018}.
These issues are not new, and similar discussions have been brewing in IR for a while.
The landmark study by Armstrong et al.~\cite{Armstrong:2009} in 2009 found that comparisons to weak baselines pervade the literature.
A decade later, is this still the case?

We began by conducting a meta-analysis to rigorously examine Lin's criticism.
His argument specifically focused on document ranking models that could be trained with commonly-available evaluation resources; specifically, such models should not require behavioral log data.
As he argued, the test collection from the TREC 2004 Robust Track (Robust04 for short) is the best exemplar of such data.
In order to restrict the scope of our meta-analysis, we followed this line of reasoning and compiled a list of all papers that have reported experimental results on Robust04.

We exhaustively examined every publication from 2005 to 2018 in the following venues to identify those that reported results on Robust04:\ SIGIR, CIKM, WWW, ICTIR,
ECIR, KDD, WSDM, TOIS, IRJ, IPM, and JASIST.
This was supplemented by Google Scholar searches to identify a few additional papers not in the venues indicated above.
Our meta-analysis was conducted in January 2019, but after the paper acceptance we included a few more papers.
A number of exclusion criteria were applied, best characterized as discarding corner cases---for example, papers that only used a subset of the topics or papers that had metrics plotted in a graph.
In total, we examined 130 papers; of these, 109 papers contained extractable average precision values that formed the basis of the results reported below.
Note that some papers did not report AP, and thus were excluded from consideration.
All papers and associated data are publicly available for verification and further analysis.\footnote{\url{https://github.com/lintool/robust04-analysis}}

For each of the 109 papers, we extracted the highest average precision score achieved on Robust04 by the authors' proposed methods, regardless of experimental condition (ignoring oracle conditions and other unrealistic setups). 
We further categorized the papers into neural (18) and non-neural (91) approaches.
Methods that used word embeddings but not neural networks directly in ranking were considered ``neural'' in our classification.
From each paper we also extracted the authors' baseline:\ in most cases, these were explicitly defined; if multiple were presented, we selected the best.
If the paper did not explicitly mention a baseline, we selected the best comparison condition using a method {\it not} by the authors (or based on previous work).

A visualization of our meta-analysis is presented in Figure~\ref{fig:scatter}.
For each paper, we show the baseline and the best result as an empty circle and a filled circle (respectively), connected by a line.
All papers are grouped by their publication year.
Neural approaches are shown in blue, and non-neural approaches in red.
We also show two regression trendlines, for non-neural (red) as well as neural approaches (blue).
A number of reference conditions are plotted as horizontal lines:\
the best submitted run at the TREC 2004 Robust Track (TREC best) at 0.333 AP is shown as a solid black line, and the median TREC run under the ``title'' condition at 0.258 AP is shown as a dotted black line (TREC median).
Finally, we show the effectiveness of an untuned RM3 run (i.e., default parameters) from the Anserini system (see Section~\ref{section:add}).


Our meta-analysis shows that researchers still frequently compare against weak baselines:\
In 36 papers (33\%), the baseline was below the TREC median.
In fact, 25 papers (23\%) reported best results that are below the TREC median and 65 papers (60\%) reported best results that are below untuned RM3 in Anserini.
Across all 109 papers, only 6 (5.5\%) reported scores higher than the TREC best.
The highest AP we encountered was by Cormack et al.~\cite{Cormack_2009} in 2009, at 0.3686.
Across over a decade's worth of publications, we see no obvious upward trend in terms of effectiveness.

Focusing specifically on the neural approaches, 8 out of 18 papers (44\%) used a baseline that is below the TREC median;
in fact, 4 papers (22\%) reported best results that were still below the TREC median.
The best results in most papers (12 or 67\%) are still below untuned RM3 in Anserini.
The highest reported scores we encountered were 0.3278 AP reported by Yang et al.~\cite{Victor2019} and 0.5381 nDCG@20 reported by MacAvaney et al.~\cite{MacAvaney2019} (the authors did not report AP results and hence the paper was excluded from the 18).
Only recently have neural models beat Lin's baselines, and the best neural models still remain quite a bit worse than the best non-neural models in terms of AP.

It is noted that not all papers purport to advance retrieval effectiveness (for example, papers about efficiency, proposing different frameworks, etc.).
Nevertheless, we believe that our visualization provides an accurate high-level snapshot of the state of the field on this test collection.
It appears that Lin's admonishments about continued use of weak baselines and skepticism about neural ranking models are warranted.

\section{Examining Additivity}
\label{section:add}

Beyond revealing comparisons to weak baselines as widespread, Armstrong et al.~\cite{Armstrong:2009} further examined why exactly this was methodologically problematic.
Such comparisons lead to improvements that ``don't add up'' because of non-additive gains.
The prototypical research paper on {\it ad hoc} retrieval proposes an innovation and compares it to a baseline that does not include the innovation; as expected, the innovation leads to increases in effectiveness.
In this way, researchers collectively introduce dozens of different innovations, all of which improve on their respective baselines.

The key question, however, is whether the effectiveness gains of these innovations are additive.
This might not occur, for example, if they exploit the same relevance signals.
To put more precisely, does an improvement over a weak baseline still hold if applied to a strong baseline?
If the answer is {\it no}, then gains over weak baselines may be illusory, and from a methodological perspective, we should not accept gains as ``real'' and ``meaningful'' unless they improve over strong baselines.
Armstrong et al.~\cite{Armstrong:2009} presented some evidence that many improvements are not additive, a finding which has been confirmed and expanded on by Kharazmi et al.~\cite{Kharazmi_etal_TOIS2016}.
However, the debate is not fully settled, as Akcay et al.~\cite{Akcay:2017:AWB:3121050.3121059} demonstrated additivity in search result diversification after better parameter tuning.

In the second part of our study, we explicitly examine the additivity hypothesis with respect to recent neural ranking models.
Specifically, we applied neural ranking models on top of the strong baselines that Lin used to make his arguments, which showed that a well-tuned implementation of query expansion based on RM3~\cite{Abdul-Jaleel04} beats the average precision reported in two recent neural IR papers, anonymously referred to as ``Paper 1'' and ``Paper 2''.

\subsection{Experimental Setup}

We began by replicating Lin's results with the Anserini toolkit~\cite{Yang_etal_JDIQ2018}, using exactly the same experimental settings (tokenization, stemming, etc.)\ described in an online guide.\footnote{\url{http://anserini.io}}
These runs used exactly the same cross-validation splits as Paper 1 (two-fold) and Paper 2 (five-fold), thus supporting a fair comparison.

On top of Lin's runs, we applied a number of neural ranking models from MatchZoo (version 1.0)~\cite{fan2017matchzoo}:\
DSSM~\cite{huang2013learning}, CDSSM~\cite{shen2014learning}, DRMM~\cite{guo2016deep}, KNRM~\cite{xiong2017end}, DUET~\cite{mitra2017learning}.
These models were selected because they were specifically designed for {\it ad hoc} retrieval;
other models available in MatchZoo, such as ARC-I~\cite{hu2014convolutional}, MV-LSTM~\cite{wan2016deep}, and aNMM~\cite{yang2016anmm} were mainly designed for short texts and not geared towards handling documents (which are much longer).
MatchZoo is implemented in Keras, using the TensorFlow backend.

The neural models were deployed in a reranking setup, where the output of the models were linearly interpolated with scores from the RM3 baseline: $\textrm{score} = \alpha \cdot \textrm{score}_{\textrm{{\tiny NN}}} + (1 - \alpha) \cdot \textrm{score}_{\textrm{{\tiny RM3}}}$.
Note that this design allows the possibility of disregarding the RM3 scores completely, with $\alpha=1$.
In our architecture, Anserini passes the raw text (minus markup tags) of the retrieved documents to MatchZoo, which internally handles document processing (tokenization, embedding lookup, etc.) prior to inference.


Following established practice, all models were trained using only the documents in the baseline RM3 runs that appear in the Robust04 relevance judgments.
We used word vectors pre-trained on the Google News corpus (3 billion words).
The entire test collection has 249 topics (with relevance judgments).
For the two-fold cross-validation condition to match Paper~1, we randomly sampled 25 topics from the training fold as the validation set; the other fold serves as the test set.
For the five-fold cross-validation condition to match Paper~2, we selected three folds for training, one fold for validation, and used the remaining fold for testing.
In all cases, we selected model parameters to maximize average precision on the development test.
The weight $\alpha$ for score interpolation with RM3 was selected in the same manner.
We set the maximum training epochs to five and used early stopping with five patience iterations.
The batch size was set to 100 and all ``title'' queries were padded to ten tokens.
Other hyperparameters were tuned using the validation set.
All models were trained on an NVIDIA GeForce GTX 1080 GPU; it takes about one minute to train the DRMM model and a few hours for the others.

\begin{table}[t]
\vspace{0.2cm}
\centering
\begin{tabular}{lll}
\toprule
\textbf{Condition} \mbox{\hspace{0.5cm}} & \textbf{AP} \mbox{\hspace{1.0cm}} & \textbf{NDCG@20} \\ 
\toprule
BM25~\cite{guo2016deep} & 0.255 & 0.418 \\
DRMM~\cite{guo2016deep} & 0.279 & 0.431 \\
\midrule
\multicolumn{3}{l}{{\it 2-fold results from Paper 1}} \\
Paper 1    & 0.2971        & - \\
BM25+RM3   & 0.2987        & 0.4398     \\
~+ DSSM    & 0.2993      & 0.4467     \\  
~+ CDSSM   & 0.2988      & 0.4455     \\
~+ DRMM    & 0.3126$^{\dagger}$  & 0.4646$^{\dagger}$ \\  
~+ KNRM    & 0.3033      & 0.4423     \\  
~+ DUET    & 0.3021      & 0.4471     \\ 
\midrule
\multicolumn{3}{l}{{\it 5-fold results from Paper 2}} \\
Paper 2    & 0.272       & - \\
BM25+RM3   & 0.3033      & 0.4514     \\
~+ DSSM    & 0.3026      & 0.4491     \\  
~+ CDSSM   & 0.2995      & 0.4468     \\
~+ DRMM    & 0.3152$^{\dagger}$  & 0.4718$^{\dagger}$ \\  
~+ KNRM    & 0.3036      & 0.4441     \\  
~+ DUET    & 0.3051      & 0.4502     \\ 
\bottomrule
\end{tabular}
\vspace{0.2cm}
\caption{Experimental results applying neural models to rerank a strong baseline; $^{\dagger}$ indicates statistical significance.}
\label{tab:exp-robust04}
\vspace{-0.6cm}
\end{table}

\subsection{Results}

Our experimental results are shown in Table~\ref{tab:exp-robust04}.
Of all the neural models we examined in MatchZoo, only the original DRMM paper evaluated on Robust04;
the first two rows show the DRMM results and their BM25 baseline (both copied from the original paper~\cite{guo2016deep}).
The paper reported a fairly substantial gain in AP, but based on our meta-analysis, the baseline is right around the TREC median and the DRMM score is still below Anserini RM3.

The second and third blocks of Table~\ref{tab:exp-robust04} report results from the two-fold and five-fold cross-validation conditions to match Paper~1 and Paper~2.
Results from Paper 1 and Paper 2 are provided for reference (neither report NDCG@20).
Note that our BM25+RM3 results are slightly higher than the results reported by Lin~\cite{Lin_SIGIRForum2018} because of code improvements after the publication of the article.
We see that our ``baseline'' already beats the best results reported in Paper 1 and Paper 2.
Based on our meta-analysis, an AP score of 0.3033 (five-fold) beats 86 out of 109 papers (79\%) and all but two neural models.

Experiments show that reranking our strong baseline with neural models yields small improvements in many cases.\footnote{The reader might wonder how it is possible that a neural model actually makes results worse, since a setting of $\alpha=1.0$ would ignore the neural model scores. However, due to cross-validation, this may not be the learned parameter.}
Statistical significance of metric differences was assessed using a paired two-tailed {\it t}-test:\ the only observed significant difference is with DRMM ($p = 0.0032$).
Even correcting for multiple hypothesis testing (e.g., Bonferroni correction), this difference remains statistically significant.
Our five-fold cross-validation result of 0.3152 with DRMM beats 98 out of 109 papers (90\%) and all but one neural model; while this can certainly be characterized as a competitive result based on our meta-analysis, it is still quite far from the best known result on Robust04 (0.3686 AP).

\subsection{Discussion}

We specifically tackle a number of shortcomings and limitations of our study.
First, only the five models implemented in MatchZoo were examined, and the quality of those implementations might be questioned.
We concede this point, and so our findings apply to only the {\it MatchZoo implementations} of the various neural models.
Nevertheless, MatchZoo has gained broad acceptance in the community as a solid experimental platform on which to explore neural ranking tasks.

The next obvious objection is that we've only examined these particular five neural ranking models.
This, of course, is valid criticism, but an exhaustive study of {\it all} models would be impractical.
We argue that the models selected are representative of the types of approaches pursued by researchers today, and that these results suffice to support at least some tentative generalizations.

The next criticism we anticipate concerns our evidence combination method, simple linear interpolation of scores.
While there are much more sophisticated approaches to integrating multiple relevance signals, this approach is commonly used~\cite{zamani2016embedding,ganguly2015word,gysel2018neural,yang2019end,rao2018multi}.
In a separate experiment where we explicitly ignored the retrieval scores, effectiveness was significantly lower.
We leave open the possibility of better evidence aggregation methods, but such future work does not detract from our findings here.

Another possible criticism of our study is the limited data condition, since we are training with only TREC judgments.
Surely, the plethora of training data that comes from behavioral logs must be considered.
While we do not dispute the effectiveness of neural approaches given large amounts of data, exploring the range of data conditions under which those models work is itself interesting.
We note a stark contrast here:\ for NLP tasks, researchers have been able to extract gains from neural approaches with only ``modest'' amounts of data (as a rough definition, datasets that can be created outside an industrial context without behavioral logs).
If it is the case that IR researchers cannot demonstrate gains {\it except with data only available to large companies}---this in itself would be an interesting statement about neural IR.
Mitra and Craswell~\cite{MitraBhaskar_Craswell_2017} classified DRMM as a lexical matching modeling (in fact, the model explicitly captures {\it tf} and {\it idf}).
DUET is a hybrid lexical/semantic matching model, while the others are semantic matching primarily.
One possible interpretation of our findings is that TREC judgments alone are not sufficient to train semantic matching models.

Finally, there is a modeling decision worth discussing:\
In our experiments, all models except for DRMM truncate the length of the input document to the first $K$ tokens (the \texttt{text2\_maxlen} parameter in MatchZoo).
Somewhat surprisingly, this is a practical issue that does not appear to be discussed in previous papers, but has a direct impact on model training time.
We performed a coarse-grained sweep of the parameter and discovered that a value of $K$ above 200 appears to be sufficient and doesn't seem to alter effectiveness substantially (one contributing factor might be the writing style of news articles).
The results reported here use a $K$ value of 500, which is longer than most documents, but still yields reasonable model training times.
We believe that document truncation can be ruled out as a reason why four of the five neural ranking models do not yield additive improvements.

\section{Conclusions}

We believe that our study supports the following conclusions:
At least with respect to the Robust04 test collection, it does not appear that the IR community as a whole has heeded the admonishments of Armstrong et al.~\cite{Armstrong:2009} from a decade ago.
Our meta-analysis shows that comparisons to weak baselines still pervade the literature.
The high water mark on Robust04 in terms of average precision was actually set in 2009, and no reported results since then (neural or otherwise) come close.
Focusing specifically on five neural ranking models in MatchZoo, we find that only one is able to significantly improve upon a well-tuned RM3 run in a reranking setup on this collection.
That is, at least under this limited data scenario, effectiveness gains from most neural ranking models do not appear to be additive.
While neural networks no doubt represent an exciting direction in information retrieval, we believe that at least some of the gains reported in the literature are illusory.

\smallskip \noindent {\bf Acknowledgments.} This work was
supported in part by the Natural Sciences and Engineering Research
Council (NSERC) of Canada.


\begin{thebibliography}{26}


\ifx \showCODEN    \undefined \def \showCODEN     #1{\unskip}     \fi
\ifx \showDOI      \undefined \def \showDOI       #1{#1}\fi
\ifx \showISBNx    \undefined \def \showISBNx     #1{\unskip}     \fi
\ifx \showISBNxiii \undefined \def \showISBNxiii  #1{\unskip}     \fi
\ifx \showISSN     \undefined \def \showISSN      #1{\unskip}     \fi
\ifx \showLCCN     \undefined \def \showLCCN      #1{\unskip}     \fi
\ifx \shownote     \undefined \def \shownote      #1{#1}          \fi
\ifx \showarticletitle \undefined \def \showarticletitle #1{#1}   \fi
\ifx \showURL      \undefined \def \showURL       {\relax}        \fi
\providecommand\bibfield[2]{#2}
\providecommand\bibinfo[2]{#2}
\providecommand\natexlab[1]{#1}
\providecommand\showeprint[2][]{arXiv:#2}

\bibitem[\protect\citeauthoryear{{Abdul-Jaleel et al.}}{{Abdul-Jaleel et
  al.}}{2004}]%
        {Abdul-Jaleel04}
\bibfield{author}{\bibinfo{person}{{Abdul-Jaleel et al.}}}
  \bibinfo{year}{2004}\natexlab{}.
\newblock \showarticletitle{{UMass} at {TREC} 2004: {Novelty} and {HARD}}.
  \bibinfo{booktitle}{\emph{TREC}}.
\newblock


\bibitem[\protect\citeauthoryear{{Akcay et al.}}{{Akcay et al.}}{2017}]%
        {Akcay:2017:AWB:3121050.3121059}
\bibfield{author}{\bibinfo{person}{{Akcay et al.}}}
  \bibinfo{year}{2017}\natexlab{}.
\newblock \showarticletitle{On the Additivity and Weak Baselines for Search
  Result Diversification Research}. \bibinfo{booktitle}{\emph{ICTIR}}.
\newblock


\bibitem[\protect\citeauthoryear{{Armstrong et al.}}{{Armstrong et
  al.}}{2009}]%
        {Armstrong:2009}
\bibfield{author}{\bibinfo{person}{{Armstrong et al.}}}
  \bibinfo{year}{2009}\natexlab{}.
\newblock \showarticletitle{Improvements That Don't Add Up: Ad-hoc Retrieval
  Results Since 1998}. \bibinfo{booktitle}{\emph{CIKM}}.
\newblock


\bibitem[\protect\citeauthoryear{{Cormack et al.}}{{Cormack et al.}}{2009}]%
        {Cormack_2009}
\bibfield{author}{\bibinfo{person}{{Cormack et al.}}}
  \bibinfo{year}{2009}\natexlab{}.
\newblock \showarticletitle{Reciprocal Rank Fusion Outperforms Condorcet and
  Individual Rank Learning Methods}. \bibinfo{booktitle}{\emph{SIGIR}}.
\newblock


\bibitem[\protect\citeauthoryear{{Fan et al.}}{{Fan et al.}}{2017}]%
        {fan2017matchzoo}
\bibfield{author}{\bibinfo{person}{{Fan et al.}}}
  \bibinfo{year}{2017}\natexlab{}.
\newblock \showarticletitle{{MatchZoo}: A Toolkit for Deep Text Matching}.
\newblock \bibinfo{journal}{\emph{arXiv:1707.07270}}.
\newblock


\bibitem[\protect\citeauthoryear{{Ganguly et al.}}{{Ganguly et al.}}{2015}]%
        {ganguly2015word}
\bibfield{author}{\bibinfo{person}{{Ganguly et al.}}}
  \bibinfo{year}{2015}\natexlab{}.
\newblock \showarticletitle{Word Embedding Based Generalized Language Model for
  Information Retrieval}. \bibinfo{booktitle}{\emph{SIGIR}}.
\newblock


\bibitem[\protect\citeauthoryear{{Guo et al.}}{{Guo et al.}}{2016}]%
        {guo2016deep}
\bibfield{author}{\bibinfo{person}{{Guo et al.}}}
  \bibinfo{year}{2016}\natexlab{}.
\newblock \showarticletitle{A Deep Relevance Matching Model for Ad-hoc
  Retrieval}. \bibinfo{booktitle}{\emph{CIKM}}.
\newblock


\bibitem[\protect\citeauthoryear{{Hu et al.}}{{Hu et al.}}{2014}]%
        {hu2014convolutional}
\bibfield{author}{\bibinfo{person}{{Hu et al.}}}
  \bibinfo{year}{2014}\natexlab{}.
\newblock \showarticletitle{Convolutional Neural Network Architectures for
  Matching Natural Language Sentences}. \bibinfo{booktitle}{\emph{NIPS}}.
\newblock


\bibitem[\protect\citeauthoryear{{Huang et al.}}{{Huang et al.}}{2013}]%
        {huang2013learning}
\bibfield{author}{\bibinfo{person}{{Huang et al.}}}
  \bibinfo{year}{2013}\natexlab{}.
\newblock \showarticletitle{Learning Deep Structured Semantic Models for Web
  Search using Clickthrough Data}. \bibinfo{booktitle}{\emph{CIKM}}.
\newblock


\bibitem[\protect\citeauthoryear{{Kharazmi et al.}}{{Kharazmi et al.}}{2016}]%
        {Kharazmi_etal_TOIS2016}
\bibfield{author}{\bibinfo{person}{{Kharazmi et al.}}}
  \bibinfo{year}{2016}\natexlab{}.
\newblock \showarticletitle{Examining Additivity and Weak Baselines}.
\newblock \bibinfo{journal}{\emph{TOSI}} \bibinfo{volume}{34},
  \bibinfo{number}{4} (\bibinfo{year}{2016}), \bibinfo{pages}{Article 23}.
\newblock


\bibitem[\protect\citeauthoryear{Lin}{Lin}{2018}]%
        {Lin_SIGIRForum2018}
\bibfield{author}{\bibinfo{person}{Lin}.} \bibinfo{year}{2018}\natexlab{}.
\newblock \showarticletitle{The Neural Hype and Comparisons Against Weak
  Baselines}.
\newblock \bibinfo{journal}{\emph{SIGIR Forum}} \bibinfo{volume}{52},
  \bibinfo{number}{2} (\bibinfo{year}{2018}), \bibinfo{pages}{40--51}.
\newblock


\bibitem[\protect\citeauthoryear{Lipton and Steinhardt}{Lipton and
  Steinhardt}{2018}]%
        {Lipton:1807.03341v2:2018}
\bibfield{author}{\bibinfo{person}{Lipton} {and} \bibinfo{person}{Steinhardt}.}
  \bibinfo{year}{2018}\natexlab{}.
\newblock \showarticletitle{Troubling Trends in Machine Learning Scholarship}.
\newblock \bibinfo{journal}{\emph{arXiv:1807.03341}}.
\newblock


\bibitem[\protect\citeauthoryear{{MacAvaney et al.}}{{MacAvaney et
  al.}}{2019}]%
        {MacAvaney2019}
\bibfield{author}{\bibinfo{person}{{MacAvaney et al.}}}
  \bibinfo{year}{2019}\natexlab{}.
\newblock \showarticletitle{{CEDR}: Contextualized Embeddings for Document
  Ranking}.
\newblock \bibinfo{journal}{\emph{arXiv:1904.07094}}.
\newblock


\bibitem[\protect\citeauthoryear{Mitra and Craswell}{Mitra and
  Craswell}{2017}]%
        {MitraBhaskar_Craswell_2017}
\bibfield{author}{\bibinfo{person}{Mitra} {and} \bibinfo{person}{Craswell}.}
  \bibinfo{year}{2017}\natexlab{}.
\newblock \showarticletitle{Neural Models for Information Retrieval}.
\newblock \bibinfo{journal}{\emph{arXiv:1705.01509}}.
\newblock


\bibitem[\protect\citeauthoryear{{Mitra et al.}}{{Mitra et al.}}{2017}]%
        {mitra2017learning}
\bibfield{author}{\bibinfo{person}{{Mitra et al.}}}
  \bibinfo{year}{2017}\natexlab{}.
\newblock \showarticletitle{Learning to Match using Local and Distributed
  Representations of Text for Web Search}. \bibinfo{booktitle}{\emph{WWW}}.
\newblock


\bibitem[\protect\citeauthoryear{{Rao et al.}}{{Rao et al.}}{2019}]%
        {rao2018multi}
\bibfield{author}{\bibinfo{person}{{Rao et al.}}}
  \bibinfo{year}{2019}\natexlab{}.
\newblock \showarticletitle{Multi-Perspective Relevance Matching with
  Hierarchical \mbox{ConvNets} for Social Media Search}.
  \bibinfo{booktitle}{\emph{AAAI}}.
\newblock


\bibitem[\protect\citeauthoryear{{Sculley et al.}}{{Sculley et al.}}{2018}]%
        {Sculley_etal_2018}
\bibfield{author}{\bibinfo{person}{{Sculley et al.}}}
  \bibinfo{year}{2018}\natexlab{}.
\newblock \showarticletitle{Winner's Curse? {On} Pace, Progress, and Empirical
  Rigor}. \bibinfo{booktitle}{\emph{ICLR Workshops}}.
\newblock


\bibitem[\protect\citeauthoryear{{Shen et al.}}{{Shen et al.}}{2014}]%
        {shen2014learning}
\bibfield{author}{\bibinfo{person}{{Shen et al.}}}
  \bibinfo{year}{2014}\natexlab{}.
\newblock \showarticletitle{Learning Semantic Representations using
  Convolutional Neural Networks for Web Search}.
  \bibinfo{booktitle}{\emph{WWW}}.
\newblock


\bibitem[\protect\citeauthoryear{{Van~Gysel et al.}}{{Van~Gysel et
  al.}}{2018}]%
        {gysel2018neural}
\bibfield{author}{\bibinfo{person}{{Van~Gysel et al.}}}
  \bibinfo{year}{2018}\natexlab{}.
\newblock \showarticletitle{Neural Vector Spaces for Unsupervised Information
  Retrieval}.
\newblock \bibinfo{journal}{\emph{TOIS}} \bibinfo{volume}{36},
  \bibinfo{number}{4} (\bibinfo{year}{2018}), \bibinfo{pages}{Article 38}.
\newblock


\bibitem[\protect\citeauthoryear{{Wan et al.}}{{Wan et al.}}{2016}]%
        {wan2016deep}
\bibfield{author}{\bibinfo{person}{{Wan et al.}}}
  \bibinfo{year}{2016}\natexlab{}.
\newblock \showarticletitle{A Deep Architecture for Semantic Matching with
  Multiple Positional Sentence Representations.}.
  \bibinfo{booktitle}{\emph{AAAI}}.
\newblock


\bibitem[\protect\citeauthoryear{{Xiong et al.}}{{Xiong et al.}}{2017}]%
        {xiong2017end}
\bibfield{author}{\bibinfo{person}{{Xiong et al.}}}
  \bibinfo{year}{2017}\natexlab{}.
\newblock \showarticletitle{End-to-End Neural Ad-hoc Ranking with Kernel
  Pooling}. \bibinfo{booktitle}{\emph{SIGIR}}.
\newblock


\bibitem[\protect\citeauthoryear{{Yang et al.}}{{Yang et al.}}{2016}]%
        {yang2016anmm}
\bibfield{author}{\bibinfo{person}{{Yang et al.}}}
  \bibinfo{year}{2016}\natexlab{}.
\newblock \showarticletitle{aNMM: Ranking Short Answer Texts with
  Attention-Based Neural Matching Model}. \bibinfo{booktitle}{\emph{CIKM}}.
\newblock


\bibitem[\protect\citeauthoryear{{Yang et al.}}{{Yang et al.}}{2018}]%
        {Yang_etal_JDIQ2018}
\bibfield{author}{\bibinfo{person}{{Yang et al.}}}
  \bibinfo{year}{2018}\natexlab{}.
\newblock \showarticletitle{{Anserini}: Reproducible Ranking Baselines Using
  {Lucene}}.
\newblock \bibinfo{journal}{\emph{JDIQ}} \bibinfo{volume}{10},
  \bibinfo{number}{4} (\bibinfo{year}{2018}), \bibinfo{pages}{Article 16}.
\newblock


\bibitem[\protect\citeauthoryear{{Yang et al.}}{{Yang et al.}}{2019a}]%
        {yang2019end}
\bibfield{author}{\bibinfo{person}{{Yang et al.}}}
  \bibinfo{year}{2019}\natexlab{a}.
\newblock \showarticletitle{End-to-End Open-Domain Question Answering with
  {BERTserini}}. \bibinfo{booktitle}{\emph{NAACL}}.
\newblock


\bibitem[\protect\citeauthoryear{{Yang et al.}}{{Yang et al.}}{2019b}]%
        {Victor2019}
\bibfield{author}{\bibinfo{person}{{Yang et al.}}}
  \bibinfo{year}{2019}\natexlab{b}.
\newblock \showarticletitle{Simple Applications of {BERT} for Ad Hoc Document
  Retrieval}.
\newblock \bibinfo{journal}{\emph{arXiv:1903.10972}}.
\newblock


\bibitem[\protect\citeauthoryear{Zamani and Croft}{Zamani and Croft}{2016}]%
        {zamani2016embedding}
\bibfield{author}{\bibinfo{person}{Zamani} {and} \bibinfo{person}{Croft}.}
  \bibinfo{year}{2016}\natexlab{}.
\newblock \showarticletitle{Embedding-based query language models}.
  \bibinfo{booktitle}{\emph{ICTIR}}.
\newblock


\end{thebibliography}


\end{document}